\documentclass[sigconf]{acmart}

\AtBeginDocument{%
  }

\usepackage{graphicx}%
\usepackage{multirow}%
\usepackage{amsthm}%
\usepackage{mathrsfs}%
\usepackage[title]{appendix}%
\usepackage{xcolor}%
\usepackage{textcomp}%
\usepackage{manyfoot}%
\usepackage{booktabs}%
\usepackage{algorithm}%
\usepackage{algorithmicx}%
\usepackage{algpseudocode}%
\usepackage{listings}%

\usepackage{anyfontsize}
\usepackage{xspace}
\usepackage{xcolor}
\usepackage{colortbl}
\definecolor{pastelgreen}{RGB}{219,255,205}
 \usepackage{subcaption}
 \usepackage{siunitx}
 \usepackage{multirow}

\newcommand{\etal}[0]{\emph{et al.}\xspace}
\newcommand\ie{\emph{i.e.},\xspace}
\newcommand\eg{\emph{e.g.},\xspace}
\newcommand\wrt{\emph{w.r.t}\xspace}
\newcommand{\citesec}[1]{Section~\ref{sec:#1}\textbf{\textbf{}}}
\newcommand{\citefig}[1]{Figure~\ref{fig:#1}}

\newcommand{\citetable}[1]{Table~\ref{table:#1}}
\newcommand{\jh}{JHipster\xspace}

\newcolumntype{L}[1]{>{\raggedright\arraybackslash}p{#1}}
\newcolumntype{C}[1]{>{\centering\arraybackslash}p{#1}}
\newcolumntype{R}[1]{>{\raggedleft\arraybackslash}p{#1}}

\newcommand{\todo}[1]{\textbf{\textcolor{red}{TODO: #1}}}\newcommand\TODO\todo

\newcommand*\rot{\rotatebox{90}}
\usepackage{changepage}
\usepackage{framed}
\makeatletter
 {\par\unskip\endMakeFramed}
\makeatother

\definecolor{formalshade}{rgb}{0.93,0.93,0.93}
\definecolor{darkblue}{rgb}{0.2, 0.2, 0.2}

\newenvironment{formal}{%
  \def\FrameCommand{%
    \hspace{1pt}%
    {\color{darkblue}\vrule width 2pt}%
    {\color{formalshade}\vrule width 4pt}%
    \colorbox{formalshade}%
  }%
  \MakeFramed{\advance\hsize-\width\FrameRestore}%
  \noindent\hspace{-1pt}
  \begin{adjustwidth}{}{7pt}%
  \vspace{2pt}\vspace{2pt}%
}
{%
  \vspace{3pt}\end{adjustwidth}\endMakeFramed%
}

\newcounter{resultcounter}

\newcommand{\B}{\textbf}

\begin{document}

\title{Exploring Performance Trade-offs in JHipster}


\author[]{Edouard Guégain$^1$, Alexandre Bonvoisin$^1$, Cl\'ement Quinton$^1$, Mathieu Acher$^2$ and Romain Rouvoy$^1$}

\affiliation{\institution{$^1$University of Lille, CNRS, Inria, Centrale Lille, UMR 9189 CRIStAL \city{Lille} \postcode{F-59000} \country{France}}}

\affiliation{\institution{$^2$University of Rennes, IRISA, Inria, IUF \city{Rennes} \country{France}}}

\renewcommand{\shortauthors}{Guégain et al.}

\begin{abstract}
The performance of software systems remains a persistent concern in the field of software engineering.
While traditional metrics like binary size and execution time have long been focal points for developers, the power consumption concern has gained significant attention, adding a layer of complexity to performance evaluation.
Configurable software systems, with their potential for numerous configurations, further complicate this evaluation process.
In this experience paper, we examine the impact of configurations on performance, specifically focusing on the web stack generator \jh. 
Our goal is to understand how configuration choices within \jh influence the performance of the generated system.
We undertake an exhaustive analysis of \jh by examining its configurations and their effects on system performance. 
Additionally, we explore individual configuration options to gauge their specific influence on performance. 
Through this process, we develop a comprehensive performance model for \jh, enabling us to automate the identification of configurations that optimize specific performance metrics.
In particular, we identify configurations that demonstrate near-optimal performance across multiple indicators and report on significant correlations between configuration choices within \jh and the performance of generated systems.
\end{abstract}

\maketitle

\section{Introduction}\label{sec:introduction}

Modern software systems face stringent performance constraints, necessitating a delicate balance between maximizing user satisfaction and minimizing hosting costs, particularly in cloud deployments. 
Stakeholders seek superior user experiences, such as low response latency, while also minimizing CPU, RAM utilization, and energy consumption \cite{Bonvoisin_SANER_2024}.
Consequently, developers are incentivized to adopt technologies capable of meeting these performance goals~\cite{ournani2020reducing}. 
However, identifying the most suitable technologies for achieving these objectives poses significant challenges.
Assessing the performance of software components is a complex and error-prone process due to various influencing factors, including the runtime environment and input workloads.
Moreover, when software exhibits variability, the number of viable configurations can grow exponentially. 
Faced with such complexity, developers often default to standard settings or employ only a subset of configuration options~\cite{8469102,PEREIRA2021111044,guegain2023configuration}.
Hence, there is a need to enhance our comprehension of overall performance in configurable software systems.

This paper undertakes an exhaustive analysis of individual configuration options and their potential correlation with various systems' performances, ranging from traditional metrics like binary size and execution time, to power consumption. 
We provide a comprehensive view of the impact of such configuration choices, that stakeholders can leverage to decide \emph{(i)} which performance indicators to use to measure the system's performance, \emph{(ii)} what performance indicators hold relevance and \emph{(iii)} how can high-performance configurations be systematically designed.

This study is conducted on the web stack generator \jh~\cite{halin2019test, halin2017yo}, an open-source development platform that facilitates the creation, development, and deployment of modern web applications.
In particular, this paper assesses the performance of systems generated by \jh across a spectrum of configurations.
Indeed, \jh offers a range of options that can be combined into more than $100,000$ configurations~\cite{Munoz_SPLC_2022}. Hereafter, we use \emph{configuration} to refer to a software system generated using a specific \jh configuration.
We chose \jh as it satisfies several critical criteria, including preexistence, open-source accessibility, comprehensive documentation, automation, and the provision of diverse metrics for evaluation.
Besides aligning seamlessly with the aforementioned prerequisites, \jh also presents the advantages of being widely employed in industrial applications, and
exhibits a proven track record, backed by over $50,000$ commits from more than $700$ contributors, and exceeding $21,200$ stars on its GitHub repository.

In particular, we make the following contributions :
\begin{itemize}
    \item[-] We present our findings on the \jh configurations, investigating how the choice of options affects a set of performance metrics.
    \item[-] Based on our synthesized findings, we compute correlations over different indicators, looking for trends related to specific indicators.
    \item[-] We provide a replication package that includes the entire dataset of performance indicators, facilitating future investigations in the domain of performance assessment for configurable software.
\end{itemize}

In the remainder of this article, \citesec{related} introduces the related work \wrt performance of configurable software and \jh. 
\citesec{subject} presents the research questions and elaborates on the performance indicators that were observed and the methodology employed to monitor them. 
\citesec{perfconfig} and \citesec{perffeatures} present and discuss the results of the experiment \wrt configurations and options, respectively.
Finally, \citesec{disc} provides a critical discussion, and \citesec{conclusion} concludes this article.

\section{Related Work}\label{sec:related}

Configurable systems are subject to combinatorial explosion: a system offering $n$ boolean options can lead to the generation of $2^n$ configurations.
Their configuration space (\ie the set of valid configurations) thus tends to be too large for an exhaustive analysis, as it is not possible to generate and test all the configurations.
Understanding such configurable systems and their related performance is challenging, and is thus an active research field.

\subsection{Performance of configurable software}

Many efforts have been put into managing the performance of configurable software. 
A survey by Pereira~\etal reports on $69$ publications in that field~\cite{pereira2021}, 
and shows that research has mainly addressed one of the following three areas, discussed thereafter.

\paragraph{Performance prediction.} 
This research area aim to estimate the performance of configuration without measuring them.
Several performance prediction approaches have been proposed in the literature~\cite{Kaltenecker2020,Guo2013,siegmund2015performance}.
Such approaches aim to build performance-influence models estimating the impact of each option on a system. 
These approaches rely on machine learning techniques to infer performance data from a sample of configurations.
One of their main objectives is to detect feature interactions---as they can have a significant impact on performances---and provide more accurate predictions than approaches that do not consider such interactions.
Relying on such performance models, Siegmund~\etal predict the performance of configurations as the sum of the impact of each feature on performances~\cite{Siegmund2012,Siegmund2013}.
In particular, they empirically observed that different releases of the same system have different performances \cite{Kaltenecker_ESEM_2023}, and that performances of configurable systems have a high level of uncertainty \cite{Dorn_ESEM_2023}. Arcaini~\etal attempted to account for such uncertainty \cite{Arcaini_SPLC_2021}.

\paragraph{Performance optimization.}
Many approaches have been proposed to address performance optimization for configurable systems.
Such approaches strive to locate near-optimal configurations \wrt some performance indicators.
Several studies provide deterministic approaches~\cite{guegain2021reducing,SVOGOR201930,Olaechea2012,8469102} to tackle this challenge.
Other authors, like Hierons~\etal, rely on genetic algorithms to minimize the number of measurements needed to optimize configurations~\cite{Hierons2016}, or leverage the performance prediction methods discussed above~\cite{Siegmund2012conqueror}.
Siegmund~\etal assess different machine learning techniques for estimating energy consumption for different configurable software systems and acknowledge that machine learning alone is insufficient to optimize software performance~\cite{9734271}. 
Recently, Weber et al.~\cite{10172770} studied whether there is a correlation between runtime performance and energy consumption of software configurations.
Following a white-box approach, the authors also aim to identify relevant functions to obtain an accurate proxy for energy consumption. 
Our work aligns with their objectives, but there are notable differences in our study. For instance, we consider a wider range of workloads and metrics, and we analyse features at component level rather than source-code level. In the context of JHipster, identifying functions at the source code level presents a challenge due to the involvement of multiple technologies (e.g., REST APIs, databases) and languages (e.g., Java, SQL, DockerFiles, ...). To gain a deeper understanding, we focus on a single case study, comprehensively covering a large portion of the configuration space by employing various workloads and indicators, thereby sacrificing some external generality for depth of analysis.

\paragraph{Recommender systems.} 
In recent years, we observed an increased interest in studies applying tools and approaches to assist users during the configuration of their systems, based on their functional needs.
For instance, Pereira \etal propose a visual recommender system  \cite{pereira2016feature} based on proximity and similarity between features. 
Similarly, Zhang \etal \cite{Zhang_ICSE_2014} use dynamic profiling and analyze the stack trace of the system to locate features that can be changed without altering the functional behavior of the system.
Other approaches, such as \cite{metzger2022} or \cite{horcas2019context}, aim at updating the configuration of the system while it is running, to adapt it to the evolution of its environment.
Such approaches rely on the performance of a subset of configurations sampled from the configuration space, and extrapolate such data to the complete configuration space. 
In contrast to such work, our focus does not lie in predicting performance or sampling configurations, but rather in understanding the system's performance by conducting an exhaustive assessment.
To the best of our knowledge, only a limited number of publications have conducted a comprehensive evaluation of an entire configuration space.
In particular, exhaustive assessments in these works primarily serve to validate prediction or optimization approaches~\cite{kolesnikov2019tradeoffs, ICO, guegain2023configuration}.
However, such papers did not conduct a specific analysis of the correlations and trends within performance indicators or the performance of individual features.

\subsection{\jh}
\jh has already been studied in previous research work, making its variability model partly available.
In particular, Halin~\etal were the first to consider \jh as a relevant case study~\cite{halin2017yo}.
They proposed to build a {\em Software Product Line} (SPL) to manage the variability of \jh, in order to leverage existing tools and approaches from the SPL community on this industrial project. 
Then, in~\cite{halin2019test}, they compare how effective different sampling approaches are in detecting defects in the source code of \jh.

The variability of \jh has also been leveraged to assess analysis methods and feature models reasoning approaches.
In particular, Horcas~\etal explore the use of Monte Carlo Tree Search for analysis of feature models and compare the efficiency of different Monte Carlo algorithms~\cite{horcas2021monte}, and validate their approach on \jh.
Such work was not focused on the performance or optimization of configuration. 
Only Munoz~\etal attempted to assess and optimize configurations of \jh~\cite{Munoz_SPLC_2022}.
They propose an extension of feature models to support quality-aware reasoning for edge computing and IoT, and provide tooling support to analyze and optimize such systems.
In particular, \jh is one of the case studies assessed against five quality attributes, namely energy footprint, time, dependency, usability, and security.
However, no in-depth analysis of the performance impact of each attribute on each configuration is provided, as such data is only used to validate the optimization approach. 
Also, no conclusions are drawn from attribute correlations.

\begin{figure*}[ht!]
    \centering    
    \includegraphics[width=0.9\textwidth]{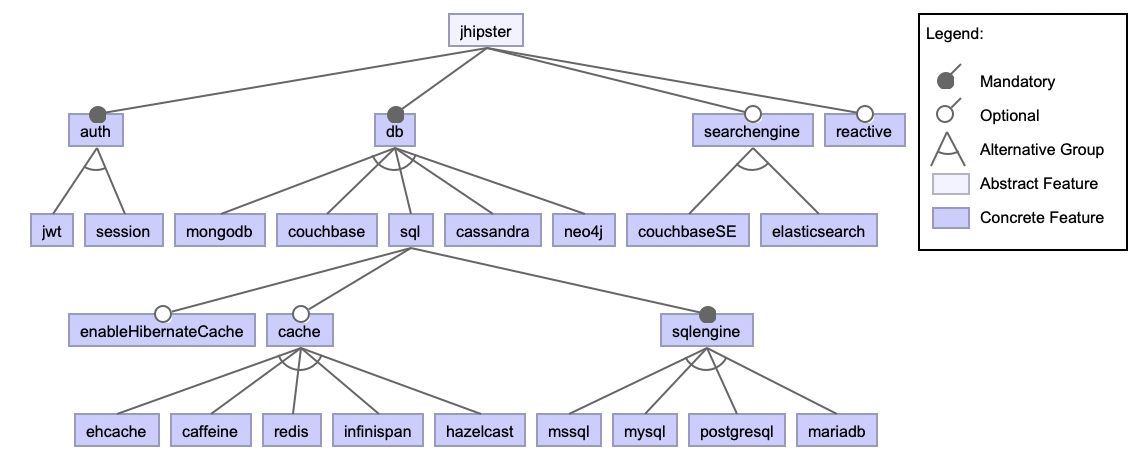}
    \caption{The variability of \jh explored in this paper. Cross-tree constraints are hidden.}
    \label{fig:model}
\end{figure*}

\section{Methodology \& Setup}\label{sec:subject}

This paper investigates the performance of a real-world highly-configurable software, regarding various performance indicators.
In particular, we aim to answer the following research questions:\\

\textbf{RQ\,1}: \textit{How are the different performance indicators related?}
Variability is known to impact configuration performances~\cite{pereira2021}. However,
collecting a large set of performance indicators is a tedious and error-prone process.
Thus, if some performance indicators are strongly correlated, and if this correlation is known and understood, the collection of some performance indicators may prove unnecessary.

\textbf{RQ\,2}: \textit{Do the different options of \jh impact the performance of generated systems in different ways?}
\jh proposes different technologies to meet technical requirements.
Setting such options may result in different performances.
Thus, analyzing the performance of configurations \wrt options they contain may highlight how such options can impact performances.

\textbf{RQ\,3}: \textit{Is it possible to drive the selection of high-performance configurations based on the performance of individual features of \jh?}
If different options of \jh exhibit different performances, then developers can leverage such knowledge to 
create highly efficient systems.

This paper does not exhaustively test all $100,000$ \jh configurations, as our analysis consists in assessing the performance of the software stack in production.
That is, our evaluation excludes options related to \emph{(i)}~building tools, such as Maven or Gradle, \emph{(ii)}~development databases (H2disk, H2mem, etc.), and \emph{(iii)}~virtualization and orchestration tools, as we assume such options do not relate to nor impact software performance in production.
As depicted by \citefig{model}, the \jh variability under study covers options related to the database system, the cache system, the search engine, the authentication method, and the usage of reactive processing.
This feature model was built based on the official documentation of \jh, by extending and updating previously published \jh feature models and through source-code mining on the \jh GitHub repository.\footnote{\url{https://github.com/jhipster}}
This model defines $118$ valid configurations, which were all run and tested to ensure that they actually work and that the selected options are properly included.
The configurations were built using version $7.9.3$ of \jh.

To answer our research questions, we empirically assessed the performances of all $118$ \jh configurations.
This section introduces the performance indicators that were monitored and the experimental protocol we followed to yield our results.
All the performance data that were monitored during our experiments are publicly available.\footnote{\url{https://zenodo.org/records/8140600}}

\begin{table*}
\centering
\renewcommand{\arraystretch}{1}
\begin{tabular}{lr|rrrrrrrrl}
\multicolumn{1}{r}{\textbf{Indicator}} &
  \multicolumn{1}{r}{\textbf{Unit}} &
  \multicolumn{1}{r}{\textbf{mean}} &
  \multicolumn{1}{r}{\textbf{std}} &
  \multicolumn{1}{r}{\textbf{min}} &
  \multicolumn{1}{r}{\textbf{25\%}} &
  \multicolumn{1}{r}{\textbf{50\%}} &
  \multicolumn{1}{r}{\textbf{75\%}} &
  \multicolumn{1}{r}{\textbf{max}}  &
  \multicolumn{1}{r}{\textbf{factor}}&
  \multicolumn{1}{l}{\textbf{Description}} \\
\textbf{size}            & Mo           & 1364.71 & 605.44 & 689.00 & 766.25 & 1373.50 & 1687.00 & 2719.00 & 3.95 & Total size of all containers \\
\textbf{services}        & $\varnothing$ & 2.63    & 0.61   & 2.00   & 2.00   & 3.00    & 3.00    & 4.00    & 2.00  & Number of containers\\
\textbf{boot-time}       & s            & 12.96   & 11.85  & 3.53   & 5.78   & 6.98    & 9.80    & 37.68   & 10.68 & Delay before stack is operational\\
\textbf{features}        & $\varnothing$  & 3.95    & 0.99   & 2.00   & 3.00   & 4.00    & 5.00    & 5.00    & 2.50 & Number of selected features\\
\hline
\textbf{auth.tion}  & ms           & 82.36   & 5.33   & 76.00  & 79.00  & 81.00   & 85.00   & 104.00  & 1.37  & User authentication\\
\textbf{auth.ted}   & ms           & 7.98    & 3.68   & 5.00   & 6.00   & 7.00    & 8.00    & 25.00   & 5.00 & First authenticated request \\
\textbf{getall}          & ms           & 11.07   & 5.56   & 6.00   & 7.00   & 8.00    & 14.00   & 23.00   & 3.83 & Fetching all objects of an entity \\
\textbf{create}          & ms           & 13.59   & 5.20   & 7.00   & 9.00   & 13.50   & 18.00   & 29.00   & 4.14  & Creating an object\\
\textbf{get}             & ms           & 6.91    & 2.41   & 5.00   & 6.00   & 6.00    & 7.00    & 20.00   & 4.00 & Fetching the created object \\
\textbf{delete}          & ms           & 16.53   & 8.84   & 5.00   & 8.00   & 12.50   & 25.75   & 29.00   & 5.80  & Deleting the creating object \\
\hline
\textbf{idle-cpu}        & W            & 1.34    & 1.55   & 0.16   & 0.69   & 0.96    & 1.54    & 9.22    & 59.00 & CPU when booted and idle\\
\textbf{idle-ram}        & W            & 0.80    & 0.78   & 0.09   & 0.43   & 0.63    & 1.01    & 4.54    & 52.59 &  RAM when booted and idle\\
\textbf{idle-total}      & W            & 2.14    & 2.32   & 0.27   & 1.10   & 1.54    & 2.52    & 13.60   & 50.96 &  CPU \& RAM when booted and idle\\
\textbf{wl-cpu}          & W            & 4.70    & 1.10   & 3.30   & 4.08   & 4.44    & 5.07    & 8.94    & 2.71  &  CPU under load \\
\textbf{wl-ram}          & W            & 2.31    & 0.63   & 1.56   & 1.98   & 2.27    & 2.60    & 4.99    & 3.21 &  RAM under load \\
\textbf{wl-total}        & W            & 7.01    & 1.72   & 4.86   & 6.14   & 6.76    & 7.67    & 13.95   & 2.87 &  CPU \& RAM under load \\
\end{tabular}%
\caption{Overview of the measured performance indicators.}
\label{table:datasummary}
\end{table*}

\subsection{Measured performance indicators}
The configurations were analyzed \wrt a set of performance indicators related to response time, energy consumption, and various static metrics were measured and analyzed. To ensure realistic simulations, response time and power usage are not limited to a single metric. Instead, multiple indicators that cover different usage scenarios of JHipster were considered. This approach allowed for evaluating JHipster's performance under various conditions and workloads.

\paragraph{Response time}
Response times were measured through Gatling, a benchmarking tool designed to simulate heavy loads on web applications and measure their performance and stability.
In particular, we monitored requests, such as the authentication request (\texttt{auth.tion}), the first authenticated request (\texttt{auth.ted}), the retrieval of all objects belonging to a certain entity (\texttt{getall}), the creation of an object of a given entity (\texttt{create}), the retrieval of this object (\texttt{get}), and the deletion of this object (\texttt{delete}).
For each indicator, we measured the mean response time.

\paragraph{Power usage}
To account for execution time variations between configurations, the energy efficiency of a configuration is defined as a consumption rate, \ie power usage, expressed in Watts.
The power usage is captured during a fixed idle period \texttt{idle-total} once the stack is booted, and then during the Gatling workload, reported as \texttt{wl-total}.
To increase the granularity of power indicators, the total power usage is further distributed into two indicators, CPU power usage, and RAM power usage.
Thus, six indicators are monitored: \texttt{idle-total}, \texttt{idle-cpu}, \texttt{idle-ram}, \texttt{wl-total}, \texttt{wl-cpu}, and \texttt{wl-cpu}.
\paragraph{Static indicators}
Some \jh configurations include external services, such as Redis or Elasticsearch, which increase the number of concurrently running containers.
Thus, the number of Docker images \texttt{service} and their \texttt{size} evolve with configurations, which may impact other performance indicators.
The number of selected features \texttt{features}, which is different from the number of images, is also monitored.
Finally, the stack's boot time \texttt{boot-time} is also accounted for in this category, as it is independent of the workload.

\subsection{Experimental setup}
To ensure reproducibility, all experiments were conducted in the same technical setup.
The experiments were performed on the Grid'5000 grid, on a single-socket server equipped with a 2.20\,GHz CPU and 96\,GB of memory, running the operating system Ubuntu 20-04 minimal.
The entire server was allocated for this experiment, preventing alterations in performances caused by other users.
The specific version of each software component is listed in our open data package.
All valid configurations were built as docker images and stored on a repository for reuse.

The assessment of each configuration is automatically performed by the following process:
First, the idle power usage of the device is monitored over a 30-second period.
Then, the generated stack is started, and the startup duration is logged.
Once the stack is ready to accept requests, the power usage is monitored over a 30-second period to compute the \texttt{idle-total} indicator.
Then, the stack is stressed with the load testing tool Gatling, while the power usage is monitored to compute the \texttt{wl-total} indicator.
The Gatling workload is the default scenario generated by \jh to prevent user bias.
Specifically, Gatling generated 100 users evenly distributed over a one-minute time window, with each user performing a scenario composed of 19 operations.
The response-time indicators are extracted from the reports generated by Gatling.

Energy consumption metrics are obtained using {\em Running Average Power Limit} (RAPL)~\cite{rapl_in_action}.
This tool provides the total power usage of the computing package, and the specific power usage of the DRAM and CPU~\cite{rapl_accurate_with_dram}.
All the power usages discussed in this article are cleaned of the idle power usage of the device.
The idle power usage of the device is measured before each workload to ensure the finest granularity.
All experiments were performed after a warmup of the device, to avoid heat-related variations in the power usage~\cite{wang2018potential}.
Finally, the performance of each configuration is estimated as the median value over five replications of the measurements.

\section{Performance of configurations}\label{sec:perfconfig}

\begin{figure*}[ht!]
     \centering
     \begin{subfigure}[b]{0.49\textwidth}
         \centering
         \includegraphics[width=\textwidth]{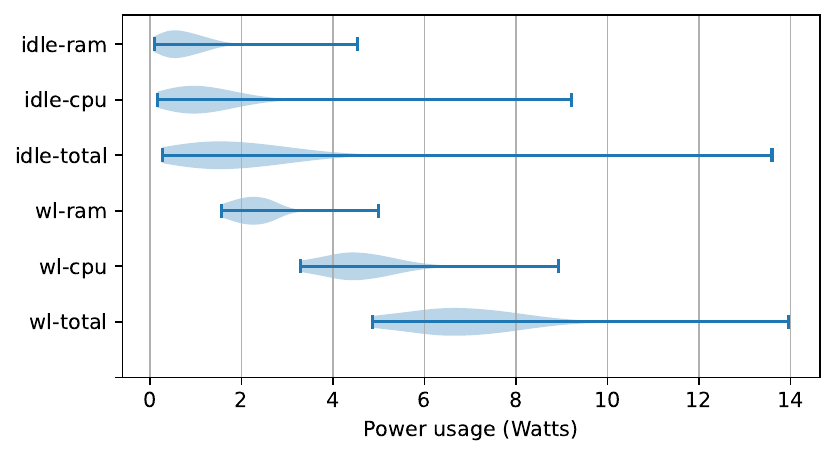}
         \caption{Variation in power usage.}
         \label{fig:stat_ener}
     \end{subfigure}
     \hfill
     \begin{subfigure}[b]{0.49\textwidth}
         \centering
         \includegraphics[width=\textwidth]{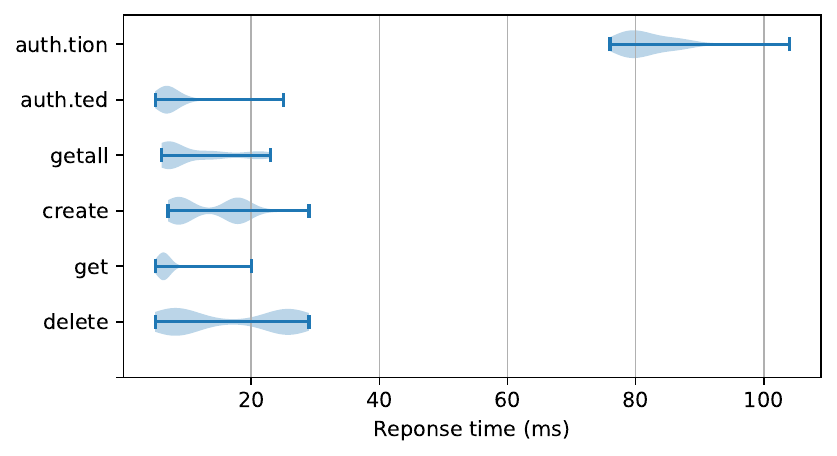}
         \caption{Variation in response time.}
         \label{fig:stat_time}
     \end{subfigure}
     \begin{subfigure}[b]{0.49\textwidth}
         \centering
         \includegraphics[width=\textwidth]{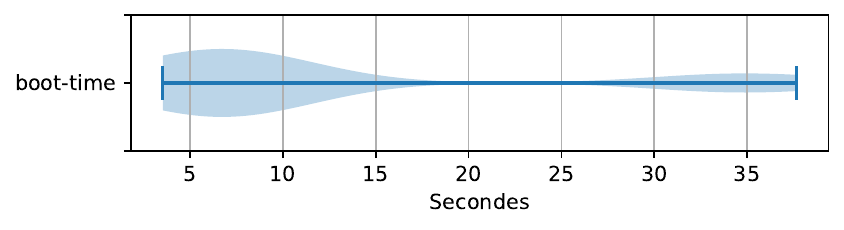}
         \caption{Variation in boot time.}
         \label{fig:stat_boot}
     \end{subfigure}
     \hfill
     \begin{subfigure}[b]{0.49\textwidth}
         \centering
         \includegraphics[width=\textwidth]{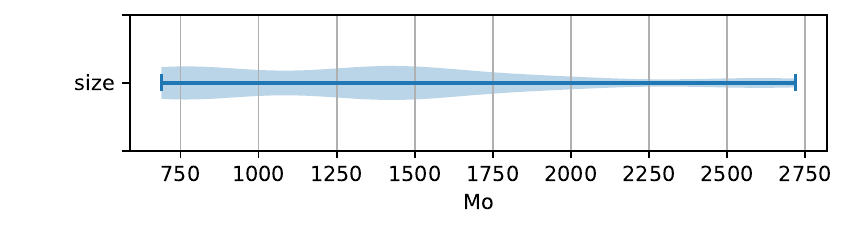}
         \caption{Variation in total size.}
         \label{fig:stat_size}
     \end{subfigure}
    \caption{Variations in the measured indicators across configurations.}
    \label{fig:all_stat_summary}
\end{figure*}

This section analyzes the performance of the $118$ configurations of \jh and look into the correlation between performance indicators.
\citetable{datasummary} depicts the variation in performances across configurations, for all the measured indicators. 
This table summarizes, for each indicator, its unit, mean, standard deviation, minimum, maximum, and quartiles. 
The factor between the maximum value and minimum value is also displayed.
While some performance indicators such as \texttt{authentication} response times are consistent across configurations, most exhibit substantial variations.
In particular, the workload power usage (\texttt{wl-total}) varies by a factor of $3$, the size by a factor of $4$, and some boot times are $11$ times longer than others. 
The total idle power usage, \ie the power usage when the configuration is running with no workload, varies by factors as high as $51$.
Therefore, one can conclude that the appropriate configuration of \jh is crucial as it can have a significant impact on the performance of the generated software system.

Beyond the difference in performances across configurations, some indicators exhibit clustering patterns, as depicted in \citefig{all_stat_summary}.
Specifically, in \citefig{stat_boot} the boot time follows a bimodal distribution: being either below $10$ seconds, or more than $30$. 
Similar behavior is visible in \citefig{stat_size}, with \texttt{size}, in \citefig{stat_time} with \texttt{create} and \texttt{delete} response times.
Finally, in \citefig{stat_ener}, most configurations are grouped around the lower values, while some outliers appear to have substantially higher power usage.

This distribution raises some questions.
Are the fastest configuration of \texttt{create} and \texttt{delete} the same configuration?
Are they also the fastest to boot?
The most power-efficient?
Such questions can be generalized to performance indicators with no obvious clusters, such as \texttt{get}.
Answering such questions requires a deeper analysis of the correlation between all the measured indicators: if some or all of these performance indicators are strongly correlated, then the assessment and optimization of performance would be substantially simplified.
Finally, it would be beneficial to identify if such behaviors are caused by specific options.

\subsection{Correlation in indicator groups}
We analyzed correlations inside each of the indicator groups---\ie between static indicators, between energetic indicators, and between response times.
The purpose of this analysis is to look for indicators that could be estimated based on other indicators.
Such indicators could act as proxies for other indicators, and thus simplify either the experimental setup or the analysis of the corresponding results.
All reported correlations in \citefig{heatmap} are Spearman Correlation coefficients. Correlations higher than 0.75 are considered as absolutely strong, and correlations lower than 0.5 are considered as absolutely weak.
These correlations are computed across the median value of the five replications, for each of the 118 configurations.

\begin{figure*}[ht!]
    \centering
    \includegraphics[width=0.65\textwidth]{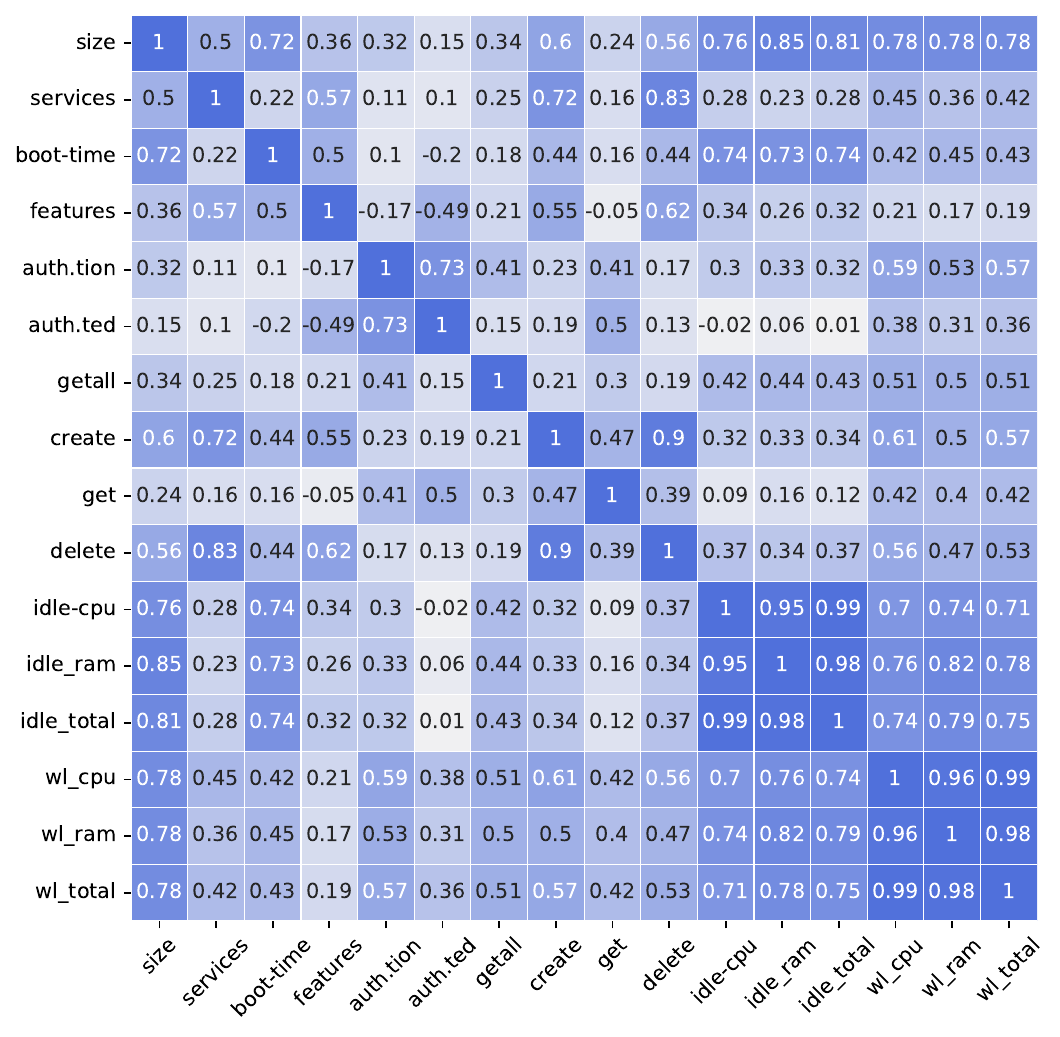}
    \caption{Correlations between indicators}
    \label{fig:heatmap}
\end{figure*}

\subsubsection{Correlation between static indicators}
Static indicators include \texttt{size}, \texttt{services}, \texttt{boot-time}, and \texttt{features}.
While all the correlations are positive, some are negligible: in particular, the correlations \texttt{boot-time} and \texttt{services} or \texttt{features} are $0.22$ and $0.5$, respectively.
By contrast, \texttt{boot-time} is strongly correlated to \texttt{size}, by $0.72$.
The indicators \texttt{services} and \texttt{features} are correlated by $0.57$.
This can be explained by the design of the variability: the services are represented as features in the feature model, but not all features are services.
Finally, \texttt{services} and \texttt{features} appear weakly correlated to \texttt{size}, by  $0.5$ and $0.36$, respectively.
Thus, a higher number of \texttt{services} or \texttt{features} tends to increase \texttt{size}, which in turn affects \texttt{boot-time}.
Therefore, \texttt{size} may be a relevant indicator to optimize, to indirectly optimize \texttt{services}, \texttt{boot-time}, and \texttt{features} metrics.

\subsubsection{Correlation between response times}
The correlations between response times indicators are all positive, suggesting that faster configurations tend to be faster in all indicators.
However, the correlations range from $0.13$ to $0.9$.
The \texttt{authentication} and \texttt{authenticated} requests are strongly correlated---\ie $0.73$.
They are only weakly correlated to other requests, such as \texttt{get} request ($0.41$ and $0.50$),\texttt{create} ($0.23$ and $0.19$) or \texttt{delete} ($0.17$ and $0.13$).
The \texttt{get} and \texttt{getall} indicators only have correlations of lower significance, below $0.5$, with the other response times.
Finally, the \texttt{create} and \texttt{delete} requests are strongly correlated ($0.9$).
Such a correlation tends to confirm the hypotheses formulated about the clustering of values in \citefig{stat_time}: the slowest configurations on \texttt{create} are also the slowest in \texttt{delete}.
The \texttt{delete} request is only weakly correlated to other response times, with a correlation lower than $0.39$.
The \texttt{create} request has weaker correlations to the other response times, between $0.3$ and $0.5$.

\subsubsection{Correlation between energetic indicators}
All the energy-related indicators are strongly correlated, with correlations ranging from $0.7$ to $1$.
The CPU and RAM power usage are strongly correlated, both in idle ($0.95$) and under load ($0.96$).
Across idle and workload, the performance of the CPU and RAM are also correlated: their respective idle and workload power usages are correlated at $0.7$ and $0.82$, respectively. Consequently, the total power usage in idle is strongly correlated to the power usage under load ($0.75$).
Since the total power usage is computed as the sum of the CPU and RAM power usage, these indicators are not independent and their correlations to the total power usage are not relevant in both idle and workload states.

\subsection{Correlations across indicators groups}
While some proxy indicators appeared inside each indicator group, assessing the performance of a configuration still requires a dedicated analysis for each indicator group.
Therefore, to further simplify performance assessment and optimization, it appears relevant to explore proxy indicators that enable the estimation of performance across different groups of indicators.

\subsubsection{Static indicators and time}
The response time to {\tt create} or {\tt delete} requests have the strongest correlation to static indicators, in particular the number of services ($0.72$ for \texttt{create} and $0.83$ for \texttt{delete}).
The correlations between \texttt{create} and \texttt{delete} and other static indicators may be the consequence of the internal correlations between static indicators.
Aside from the \texttt{create} and \texttt{delete} indicators, the response times are only weakly correlated with static indicators.
The correlations range from $-0.05$ to $-0.49$.
Thus, only \texttt{services} is a potential proxy indicator for \texttt{create} and \texttt{delete}, and static indicators do not provide a means to assess the response time of the other operations.

\subsubsection{Static indicators and power}
The correlation between power indicators and \texttt{size} ranges from $0.76$ to $0.85$.
In particular, \texttt{size} is strongly correlated to all energetic indicators (between $0.76$ and $0.85$), while \texttt{boot-time} is strongly correlated to the idle power usage (between $0.73$ and $0.74$).
The correlation between the remaining static indicators \texttt{service} and \texttt{features} and the energetic indicators are weaker, between $0.17$ and $0.45$.
However, \texttt{service} is overall more correlated to energetic under load ($0.36$ to $0.45$) than in idle ($0.23$ to $0.28$), whereas \texttt{features} exhibits the opposite behavior ($0.26$ to $0.34$ in idle, and $0.17$ to $0.21$ under load).
Therefore, \texttt{size} may be a proxy for idle and load power usages, and \texttt{boot-time} may be a proxy only for idle power usage.

\subsubsection{Time and power}
The response times are systematically more strongly correlated to the power usage under load (between $0.31$ and $0.61$) than to the one in idle (between $0.01$ and $0.44$).
The behavior may be explained by the fact that both response times and power usage underquantify the behavior of the system at runtime.
Furthermore, the power usage of the RAM is less correlated to response time than the power usage of the CPU.
The power usage under load is more correlated to  \texttt{authentication}, \texttt{getall}, \texttt{create}, and \texttt{delete} (between $0.47$ and $0.61$) than to \texttt{authenticated} and \texttt{get} (between $0.31$ and $0.42$).

\subsection{Proxy indicators}

Some of the analyzed indicators exhibit strong correlations.
In particular, all the energetic indicators are strongly correlated with each other.
The idle CPU and RAM power usage are strong predictors of the total idle power usage, and the CPU and RAM usage under load are strong predictors of the total power usage under load (with $R^2$ of respectively $0.99$, $0.98$, $0.98$, and $0.96$ in linear regression).
The total power usage in idle is also a good predictor of the total power usage under load ($R^2=0.71$).
Static indicators are also correlated with each other, and in particular \texttt{size} is a good proxy indicator for \texttt{boot-time} ($R^2=0.68$).
Proxy indicators also appear between response times, in particular with \texttt{authentication} and \texttt{authenticated} ($R^2=0.71$) and between \texttt{create} and \texttt{delete} ($R^2=0.84$).
However, proxy indicators within the same group of indicators have a limited interest. 
Indeed, except for \texttt{boot-time}, the ability to measure a given indicator of a group implies the ability to measure all indicators of that group.

Across indicator scopes, only a limited set of indicators are correlated: \texttt{size} and power usages, and some response times with \texttt{service} or the power indicators under load.
However, such correlations do not convert into strong predictors.
In particular, \texttt{size} can only approximate the total power usage under load ($R^2=0.29$). 
In idle, \texttt{size} is a better predictor of the RAM power usage than the total power usage  ($R^2$ of respectively $0.24$ and $0.16$).
\texttt{service} provides better estimates of the responses times of \texttt{create} and \texttt{delete} ($R^2$ of respectively $0.51$ and $0.68$).
Among response times, only \texttt{authentication} and \texttt{getall} allow for estimating the total power usage under load ($R^2$ of respectively $0.32$ and $0.26$). 
Other response times perform worse in such a task ($R^2<0.1$).

\begin{formal}
\textbf{RQ\,1}:
Some indicators exhibit strong correlations, in particular when they quantify measurements of similar natures, allow the use of proxy indicators to simplify performance assessment and optimization.
Such proxies can serve as warning signs that a system might exhibit poor performance in unmeasured indicators.
However, they do not allow for quantifying precisely such a performance without specific measure.
Correlations also exist across indicators of different natures, but they tend to be weaker and less frequent.
Thus, a specific evaluation of each of the indicator groups is still required.
\end{formal}

\section{Performance of options}
\label{sec:perffeatures}

This section aims to understand the impact of options of \jh on its performance and to discover means to create efficient configurations through relevant option selection.
As the previous section highlighted that the CPU, the RAM, and the total power usage are strongly correlated (more than $0.96$), only the latter is discussed in the remainder of this paper.

\subsection{Performance of individual options}

\begin{table*}
\centering
\begin{tabular}{llrrrr|rrrrrr|rr}
 &
  \multicolumn{1}{c}{\textbf{}} &
  \multicolumn{1}{c}{\rot{size}} &
  \multicolumn{1}{c}{\rot{services}} &
  \multicolumn{1}{c}{\rot{boot-time}} &
  \multicolumn{1}{c}{\rot{features}} &
  \multicolumn{1}{c}{\rot{auth.tion}} &
  \multicolumn{1}{c}{\rot{auth.ted}} &
  \multicolumn{1}{c}{\rot{getall}} &
  \multicolumn{1}{c}{\rot{create}} &
  \multicolumn{1}{c}{\rot{get}} &
  \multicolumn{1}{c}{\rot{delete}} &
  \multicolumn{1}{c}{\rot{idle-total}} &
  \multicolumn{1}{c}{\rot{wl-total}} \\

  \multirow{8}{*}{Database}        
  & \textbf{Cassandra}  & 907.00  & 3.00        & 4.05  & 2.00 & 80.00 & 8.00  & 8.00  & \cellcolor{pastelgreen}\B{7.00}  & 6.00  & \cellcolor{pastelgreen}\B{5.00}  & 1.72  & 7.37  \\
&\textbf{couchbase}     & 1730.00 & \cellcolor{pastelgreen}\B{2.00}    & 6.69  & 2.00  & 86.00 & \cellcolor{pastelgreen}\B{6.00}  & 17.00 & \cellcolor{pastelgreen}\B{7.00}  & \cellcolor{pastelgreen}\B{5.00}  & \cellcolor{pastelgreen}\B{5.00}  & 13.59 & 13.51 \\
&\textbf{mariaDB}       & 695.00  & \cellcolor{pastelgreen}\B{2.00}    & 5.13  & 2.00  & 80.00 & 8.00  & \cellcolor{pastelgreen}\B{6.00}  & 8.00  & 6.00  & 7.00  & 0.49  & \cellcolor{pastelgreen}\B{4.93}  \\
&\textbf{mongoDB}       & 740.00  & \cellcolor{pastelgreen}\B{2.00}    & \cellcolor{pastelgreen}\B{3.73}  & 2.00 & \cellcolor{pastelgreen}\B{77.00} & 7.00  & \cellcolor{pastelgreen}\B{6.00}  & 7.00  & 6.00  & 6.00  & 0.58  & 5.15  \\
&\textbf{mssql}         & 1925.00 & \cellcolor{pastelgreen}\B{2.00}    & 33.76 & 2.00  & 81.00 & 9.00  & 7.00  & 9.00  & 7.00  & 8.00  & 2.41  & 6.77  \\
&\textbf{mysql}         & 765.00  & \cellcolor{pastelgreen}\B{2.00}    & 5.39  & 2.00  & 82.00 & 10.00 & 7.00  & 9.00  & 7.00  & 9.00  & 1.07  & 6.82  \\
&\textbf{neo4j}         & 869.00  & \cellcolor{pastelgreen}\B{2.00}    & 6.68  & 2.00  & 86.00 & 12.00 & 10.00 & 12.00 & 9.00  & 9.00  & 1.19  & 6.45  \\
&\textbf{postgresql}    & \cellcolor{pastelgreen}\B{689.00} & \cellcolor{pastelgreen}\B{2.00} & 5.16  & 2.00  & 80.00 & 8.00  & 7.00  & 8.00  & 7.00  & 7.00  & \cellcolor{pastelgreen}\B{0.27}  & 5.04  \\
\hline
\multirow{2}{*}{Authentication}     
&\textbf{jwt}           & 1358.88 & 2.67 & \cellcolor{pastelgreen}\B{13.60} & 4.06 & \cellcolor{pastelgreen}\B{78.69} & \cellcolor{pastelgreen}\B{6.51}  & 10.57 & 13.29 & 6.39  & 16.98 & \cellcolor{pastelgreen}\B{1.78}  & \cellcolor{pastelgreen}\B{6.61}  \\
&\textbf{session}       & \cellcolor{pastelgreen}\B{1358.71} & 2.67 & 13.64 & 4.06 & 83.39 & 7.43  & \cellcolor{pastelgreen}\B{10.02} & \cellcolor{pastelgreen}\B{13.12} & \cellcolor{pastelgreen}\B{6.31}  & \cellcolor{pastelgreen}\B{16.86} & 1.82  & 6.66  \\
\hline

\multirow{7}{*}{Cache}              
& \textbf{No cache}     & \cellcolor{pastelgreen}\B{1353.62} & \cellcolor{pastelgreen}\B{2.50} & \cellcolor{pastelgreen}\B{12.94} & \cellcolor{pastelgreen}\B{2.50} & 82.31 & 9.19  & \cellcolor{pastelgreen}\B{6.62}  & \cellcolor{pastelgreen}\B{13.00} & 6.62  & \cellcolor{pastelgreen}\B{16.75} & \cellcolor{pastelgreen}\B{1.51}  & \cellcolor{pastelgreen}\B{6.49}  \\
&\textbf{Any cache}     & 1385.29 & 2.70 & 14.46 & 4.50 & \cellcolor{pastelgreen}\B{80.96} & \cellcolor{pastelgreen}\B{6.49}  & 11.30 & 13.46 & \cellcolor{pastelgreen}\B{6.34}  & 17.35 & 1.90  & 6.69  \\
\cmidrule{2-14}
&\textbf{caffeine}      & \cellcolor{pastelgreen}\B{1355.19} & \cellcolor{pastelgreen}\B{2.50} & 13.22 & 4.50 & 79.25 & 6.12  & 7.25  & \cellcolor{pastelgreen}\B{13.06} & 6.25  & 16.62 & \cellcolor{pastelgreen}\B{1.46}  & 6.31  \\
&\textbf{ehcache}       & 1355.38 & \cellcolor{pastelgreen}\B{2.50} & \cellcolor{pastelgreen}\B{13.10} & 4.50 & \cellcolor{pastelgreen}\B{79.06} & \cellcolor{pastelgreen}\B{6.06}  & 7.25  & \cellcolor{pastelgreen}\B{13.06} & \cellcolor{pastelgreen}\B{6.12}  & \cellcolor{pastelgreen}\B{16.38} & 1.58  & \cellcolor{pastelgreen}\B{6.29}  \\
&\textbf{hazelcast}     & 1371.31 & \cellcolor{pastelgreen}\B{2.50} & 16.10 & 4.50 & 82.31 & 6.81  & 13.62 & 13.50 & 6.50  & 17.25 & 2.96  & 7.35  \\
&\textbf{infinispan}    & 1367.88 & \cellcolor{pastelgreen}\B{2.50} & 16.36 & 4.50 & 80.25 & 6.12  & \cellcolor{pastelgreen}\B{7.19}  & 13.81 & 6.19  & 17.12 & 1.63  & 6.36  \\
&\textbf{redis}         & 1476.69 & 3.50 & 13.53 & 4.50 & 83.94 & 7.31  & 21.19 & 13.88 & 6.62  & 19.38 & 1.88  & 7.13  \\
\hline
\multirow{2}{*}{CouchbaseSE}        
&\textbf{No}            & 1733.50 & 2.00 & 6.43  & \cellcolor{pastelgreen}\B{2.50} & \cellcolor{pastelgreen}\B{86.50} & 7.00  & 19.50 & 7.00  & 5.00  & 5.00  & \cellcolor{pastelgreen}\B{13.49} & \cellcolor{pastelgreen}\B{13.34} \\
&\textbf{Yes}           & 1733.50 & 2.00 & \cellcolor{pastelgreen}\B{6.35}  & 3.50 & 87.00 & 7.00  & \cellcolor{pastelgreen}\B{19.00} & 7.00  & 5.00  & 5.00  & 13.59 & 13.85 \\
\hline
\multirow{2}{*}{Elasticsearch}      
&\textbf{No}            & \cellcolor{pastelgreen}\B{1024.48} & \cellcolor{pastelgreen}\B{2.14} & \cellcolor{pastelgreen}\B{12.84} & \cellcolor{pastelgreen}\B{3.52} & \cellcolor{pastelgreen}\B{81.96} & \cellcolor{pastelgreen}\B{7.95}  & 10.86 & \cellcolor{pastelgreen}\B{9.39}  & \cellcolor{pastelgreen}\B{6.98}  & \cellcolor{pastelgreen}\B{8.86}  & \cellcolor{pastelgreen}\B{1.45}  & \cellcolor{pastelgreen}\B{6.16}  \\
&\textbf{Yes}           & 1694.95 & 3.14 & 13.87 & 4.52 & 82.50 & 8.11  & \cellcolor{pastelgreen}\B{10.80} & 18.50 & 7.02  & 25.45 & 2.04  & 7.38  \\
\hline
\multirow{2}{*}{Reactive}          
&\textbf{No}            & \cellcolor{pastelgreen}\B{1370.08} & 2.42 & 10.42 & \cellcolor{pastelgreen}\B{2.50} & \cellcolor{pastelgreen}\B{81.08} & \cellcolor{pastelgreen}\B{8.08}  & \cellcolor{pastelgreen}\B{8.33}  & \cellcolor{pastelgreen}\B{11.83} & \cellcolor{pastelgreen}\B{6.25}  & \cellcolor{pastelgreen}\B{14.50} & 3.44  & \cellcolor{pastelgreen}\B{7.58}  \\
&\textbf{Yes}           & 1380.83 & 2.42 & \cellcolor{pastelgreen}\B{9.37}  & 3.50 & 92.42 & 16.17 & 16.83 & 17.25 & 11.67 & 14.92 & \cellcolor{pastelgreen}\B{3.24}  & 9.03 
\end{tabular}%
\caption{Average performance of configurations containing each option}
\label{table:option_perf}

\end{table*}

This section compares the performance of various options available in \jh.

\paragraph{Authentication}
In our experiment, we analyzed two authentication methods: session and JWT. 
There is no significant difference \emph{w.r.t.} energy consumption and static indicators between the two authentication methods, but some variations can be observed regarding response time.
Specifically, the authentication and first authentication requests are slower with \texttt{session} than with \texttt{JWT} by $4$\% and $12$\%, respectively.
Then, \texttt{session} is slightly faster on the remaining operations \texttt{getall} ($5$\%), and less than $2$\% faster on the remaining operations \texttt{create}, \texttt{get}, and \texttt{delete}.

\paragraph{Databases}
\jh supports $8$ databases: Mysql, Microsoft SQL Server (MSSQL), Postgres, MariaDB, Mongo, Cassandra, Neo4j, and Couchbase.
The total size of the stack differs depending on the database. 
While most database footprints are between $690$\,Mb and $900$\,Mb, the average size of configurations is $1730$\,Mb for the ones containing Couchbase and $1925$\,Mb for the ones containing MSSQL.
Thus, the choice of database appears to have a strong impact on the size of the system, and can partially explain the spread in size in \citefig{stat_size}.
Finally, while most databases have boot times between $3.7$ and $6.7$ seconds, MSSQL increases this value to $34$ seconds. 
It thus appears that MSSQL is responsible for the clustering visible in \citefig{stat_boot}.

The selected database also affect response times. For instance, Neo4j is among the slowest across all response times, except for \texttt{getall}.
Contrarily, Couchbase is among the fastest across all response times, but the slowest for \texttt{getall}.

As for energy indicators, Couchbase exhibits an outlying behavior: its power usage is higher in idle ($13,59$\,W in total) than under load ($13,51$\,W in total).
This behavior is due to the design of Couchbase, which performs opportunistic indexation in idle to optimize response times under load.
All other databases have idle power usage between $0.5$\,W and $2.4$\,W, except PostgreSQL with $0.27$\,W, and load power usage between $4.9$\,W and $6.8$\,W, except for Cassandra with $7$\,W.

\paragraph{Hibernate cache} 
\jh supports the activation of Hibernate cache for SQL databases.
This analysis only accounts for configurations where Hibernate cache is available and activated, while configurations where the Hibernate cache is available and disabled serve as a baseline.
Activating the Hibernate cache appears to increase the boot time by 12\% and the response time of \texttt{create} and \texttt{delete} requests by $3.5$\%. The response time for \texttt{getall} is increased by $71$\%.
Such overhead is leveraged in reading operations: the response times for \texttt{get} and \texttt{authenticated} are reduced by $4$\% and $29$\%, respectively.
While activating the Hibernate cache substantially increases the idle power usage ($+26$\%), it has a limited impact on the power usage under load ($+3$\%).

\paragraph{Cache provider}
When the Hibernate cache is enabled, a cache provider must be selected among the five offered by \jh.
In \citetable{option_perf}, the ``No cache" line refers to configurations where a cache system was available, but was not selected and thus acts as a baseline.
The choice of cache system affects static indicators.
Specifically, Redis increases the service count and the \texttt{size} of configurations.
Infinispan and Hazelcast increase the boot time by $26$\% and $24$\%, respectively, whereas other providers increase such boot time by less than $5$\%.

The choice of the cache system also impacts the response times.
Some operations are consistently affected. For instance, response times for \texttt{authenticated} are reduced by up to $35$\%, and \texttt{create} are increased by up to $7$\%.
However, some cache systems have outlying behavior.
For instance, Hazelcast and Redis increase \texttt{getall} by respectively $106$\% and $220$\% while other systems only increase it by $10$\%.
Similarly, Caffeine and Ehcache do not affect \texttt{delete} response times while other systems increase it by up $16$\%.
Ehcache appears to have the best performance across response times, while Caffeine is a close second.
Contrarily, Redis appears to provide the worst performance gains.

Finally, the choice of cache systems also impacts the energy efficiency of the stack.
Hazelcast and Redis substantially increase the idle power usage, by $96$\% and $25$\%, respectively 
Under load, such a difference is reduced to $+13$\% and $+10$\%, respectively.
Ehcache and Infinispan increase the idle power usage by $5$\% and $8$\%, respectively, but reduce the power usage under load by $-3$\% and $-2$\%.
Finally, Caffeine reduces the power usage by $3$\%, both in idle and under load.
Overall, configurations containing Ehcache or Caffeine exhibit the lowest power usage, both idle and under load. 

\paragraph{Search engine}
\jh supports Elasticsearch or Couchbase as search engines (S. E.), providing support to search content in the database.
Each search engine is compared to its respective baseline of configurations that could contain them but do not.
Configurations with CouchbaseSE are thus compared to configurations with Couchbase and not CouchbaseSE, while configurations with Elasticsearch are compared to configurations without Elasticsearch, Couchbase, or Cassandra.

CouchbaseSE is embedded in Couchbase, and thus does not affect \texttt{size} and \texttt{boot time}, whereas Elasticsearch increases them by $65$\% and $8$\%, respectively.
Thus, Elasticsearch can also be partially responsible for the spread in \texttt{size} visible in \citefig{stat_size}.
While both search engines have a limited impact on response times, Elasticsearch increases the response time of \texttt{create} by $97$\% and \texttt{delete} by $187$\%.
Finally, enabling the Couchbase search engine has only a limited impact on power usage: $+1$\% in idle and $+4$\% under load.
Contrarily, enabling Elasticsearch substantially increases power usage, by $41$\% in idle and $20$\% under load.

\paragraph{Reactive}
The Reactive option enables reactive programming for the application. Reactive allows for creating an asynchronous and non-blocking API, in order to improve the scalability of the system.
Reactive has a limited impact on static indicators, by reducing the boot time by $10$\%.
However, it substantially increases the response times. 
In particular, \texttt{getall} and \texttt{get} are increased by $102$\% and $87$\%, respectively.
This difference is more limited for \texttt{create}, $46$\%.
Only \texttt{delete} is unaltered.
Furthermore, reactive reduces the power usage in idle by $6$\%, but increases the power usage under load by $19$\%.

\begin{figure*}[ht!]
     \centering
     \begin{subfigure}[b]{0.49\textwidth}
         \centering
         \includegraphics[width=\textwidth]{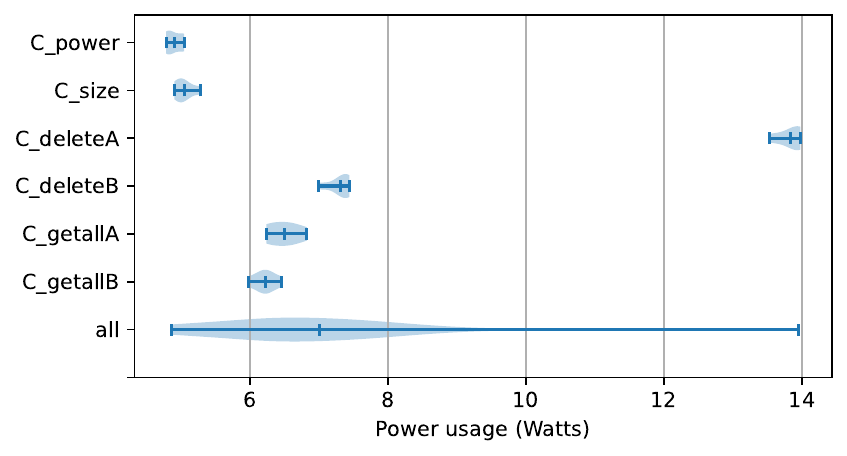}
         \caption{Variation in power usage.}
         \label{fig:opti_ener}
     \end{subfigure}
     \hfill
     \begin{subfigure}[b]{0.49\textwidth}
         \centering
         \includegraphics[width=\textwidth]{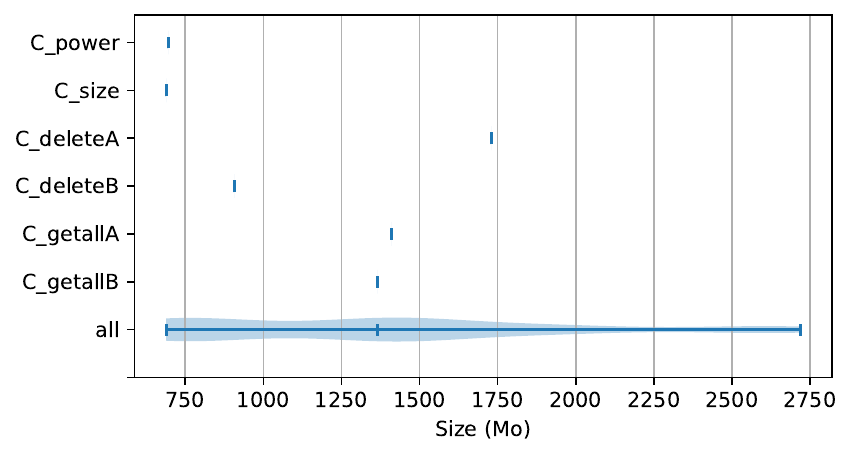}
         \caption{Variation in total size.}
         \label{fig:opti_size}
     \end{subfigure}
     \begin{subfigure}[b]{0.49\textwidth}
         \centering
         \includegraphics[width=\textwidth]{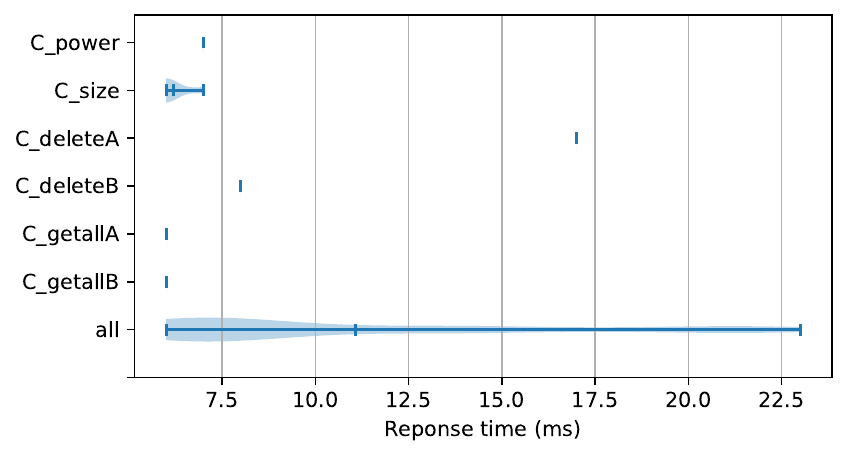}
         \caption{Variation in \texttt{getall} response time.}
         \label{fig:opti_getall}
     \end{subfigure}
     \hfill
     \begin{subfigure}[b]{0.49\textwidth}
         \centering
         \includegraphics[width=\textwidth]{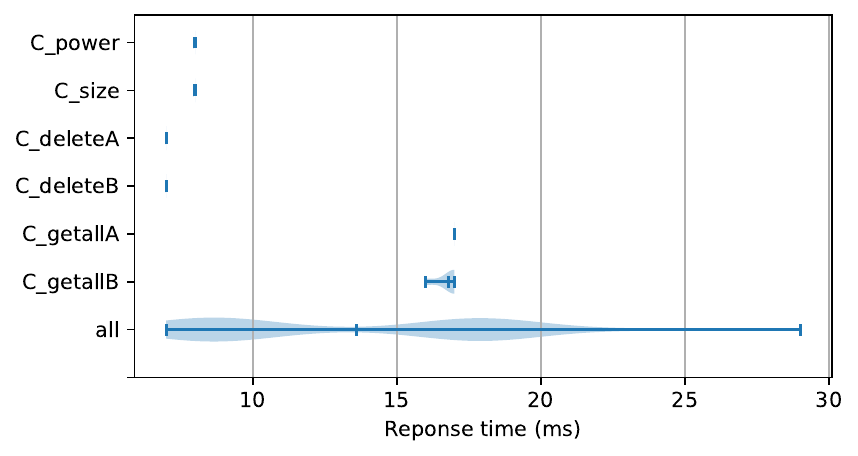}
         \caption{Variation in \texttt{delete} response time.}
         \label{fig:opti_create}
     \end{subfigure}
        \caption{Performance of the optimal configurations.}
        \label{fig:optimals}
\end{figure*}

\begin{formal}
\textbf{RQ\,2}:
The available options have different impacts on the performance of the generated system.
This impact is not consistent across indicators: the same option can have a negligible impact on a given indicator, while substantially affecting another one.
Thus, an informed feature selection per indicator is needed to achieve efficient configurations tailored to specific requirements.
\end{formal}

\subsection{Finding optimal configurations}
The previous section delivered insights into the performance of each option of \jh.
Such data can be used to compare options with each other, and thus to automate the configuration selection process by selecting  options that optimize one or many performance indicators.
However, it is unclear if such configurations are actually the best-performing \wrt their goal indicator, and how other indicators will be affected.
This section thus assesses the performance of a set of configuration, each optimizing a specific indicator, and compares such configurations with the set of all valid configurations.

To void redundancy in the validation process, we selected indicators based on the correlations found in \citesec{perfconfig}, \ie indicators that act as proxies for other indicators.
The different selected optimization goals relate to the indicators the more strongly correlated to other indicators as described in \citesec{perfconfig}---\ie the power usage during workload, the total size, and the response times for \texttt{getall} and \texttt{delete} requests. 

The creation of a configuration that optimizes a given indicator is performed by selecting options that maximize this indicator.
However, the decision tree is constrained by the feature model: the selection of an option can prevent the addition of other options later in the selection process.
To limit such constraints, the options are selected by decreasing the number of implementations: first, the database type, then the cache system if applicable, and finally the authentication method, search engine, and reactive processing.
In addition, we decided that if an optional feature (a search engine or reactive) has no impact on the targeted performance indicator, it is included in the configuration to maximize capabilities. 
Finally, the \texttt{delete} and \texttt{getall} optimization objectives generated \textit{ex aequo} configurations, which are thus differentiated as ``A" and ``B".
\citetable{bestconfigs} details the configuration for each goal.

\citefig{optimals} presents the performance of the different candidates, in each of the goal indicators.
The performance of all the valid configurations in each goal indicator is also displayed, as a reference.
With the exception of power, each candidate exhibits very strong performance on their respective goal indicator: {\tt C\_getall(A)} and {\tt C\_getall(B)} offer the best response times in \texttt{getall}, {\tt C\_delete(A)} and {\tt C\_delete(B)} offer the best response times in \texttt{create}, {\tt C\_size} has the smallest \texttt{size}, and {\tt C\_power} has the lowest power usage.
Thus, selecting relevant options during the configuration process helps improve the performance of the selected configuration.
However, this selection process needs additional formalizing and more accurate performance models to ensure optimal configurations.

\begin{table}[ht!]

\centering
\begin{tabular}{l|cccccc}
\bf Optim. Goal  & \bf Auth & \bf DB     & \bf Cache  & \bf S. E. & \bf React. \\
\hline
\tt C\_power     & JWT      & MariaDB    & Ehcache    & No            & No\\
\tt C\_size      & Session  & PostgreSQL & No         & No            & No\\
\tt C\_delete(A) & JWT      & Couchbase  & N/A         & No            & No\\
\tt C\_delete(B) & JWT      & Cassandra  & N/A         & Yes           & No\\
\tt C\_getall(A) & Session  & MariaDB    & No         & Yes           & No\\
\tt C\_getall(B) & Session  & MongoDB    & N/A         & Yes           & No\\
\end{tabular}
\caption{Best configurations per optimization goal.}
\label{table:bestconfigs}
\end{table}

While the selected goal indicators are not strongly correlated to each other, as observed in \citesec{perfconfig}, some of the different candidates have similar performances.
In particular, {\tt C\_power}, and {\tt C\_size} have both low power usage, a small size, and low response time in \texttt{getall} and \texttt{delete} compared to the complete dataset, despite having different authentication methods, database, and cache policy.
Contrarily, {\tt C\_getall(A)} and {\tt C\_getall(B)} have average performance in \texttt{power}, \texttt{size}, and \texttt{delete}.
Finally, the performance of {\tt C\_delete(A)} and {\tt C\_delete(B)} are identical in \texttt{delete} but differ substantially in \texttt{power}, \texttt{getall}, and \texttt{size}. 
In particular, {\tt C\_delete(A)} is consistently worse than {\tt C\_delete(B)} in such indicators, but also ranks among the worst configurations of the configuration space in \texttt{power}.

Thus, it appears that creating configurations composed of the best-performing options in an indicator is a relevant method to create configurations that optimize for this indicator.
However, this configuration may prove average in other indiactors (\eg {\tt C\_getall(A)} in \texttt{delete}) or even inefficient (\eg {\tt C\_delete(A)} in \texttt{getall}).
Such situations pave the way for multi-objective optimization approaches, such as~\cite{VEREL2013331}, that can be leveraged to find configurations that optimize for many objectives at once---\ie that are near-optimal in all indicators rather than optimal in a single indicator.
In this particular example, {\tt C\_power} and {\tt C\_size} are potential candidates to optimize \texttt{power}, \texttt{size}, \texttt{getall}, and \texttt{delete} as both configurations are near-optimal in all such indicators.

\begin{formal}
\textbf{RQ\,3}:
Understanding how individual features impact an indicator allows developers to identify high-performing configurations.
However, optimizing for a given indicator may compromise the overall performance of the system.
This complexity suggests the potential for multi-objective optimization strategies to identify configurations that effectively balance multiple goals. 
\end{formal}

\section{Discussion}\label{sec:disc}
As an empirical study, the results presented in this article must be interpreted \wrt to the methodology.

\textbf{Generalization.}
The results of this study can not be generalized as a ground truth about the performance of each of the analysed services, as the configurations were assessed on a specific, synthetic workload.
Furthermore, the performance model for options, built from the results of this workload, is only a performance estimation.
Such a model was only built to assess the possibility of inferring high-performance configurations for a given workload and a given performance indicator.
Finally, all the services were hosted on a single device, which may not be representative of a production deployment, in particular, \wrt to response times and power usage.

\textbf{Workload.}
The workload used to assess the configurations was automatically generated by \jh during the creation of the project.
Using this generated workload prevents bias in the selection of the workload.
However, this workload may not be representative of an actual production environment.
This affects the extrapolation of our results to other contexts and workloads, in particular, \wrt the performance of individual options discussed in \citesec{perffeatures}.
In addition, the performance was assessed over a single data model, affecting the generalization of our results to other data models~\cite{10172849}.

\textbf{Variability pruning.}
Some options offered by \jh were excluded from this analysis.
In particular, the Oracle database was excluded for licensing reasons, and micro-services or deployment options due to limitations in the benchmarking tool.
Thus, no data can be provided for such options, while they may impact performances.

\section{Conclusion}\label{sec:conclusion}

This article reports on the automated analysis conducted on the variability and performance of the configurable stack generator \jh.
This study reveals that the diverse configurations of \jh yield significantly varied performances across different performance indicators, emphasizing the impact of configuration on \jh-built system performance.
However, strong correlations were observed among certain performance indicators, particularly those quantifying measurements of a similar nature.
These correlations enable the use of proxy indicators to streamline performance assessment and optimization efforts.
While correlations also exist between indicators of different natures, they tend to be less pronounced and less frequent. 
Consequently, a dedicated performance assessment for each indicator group remains necessary.

The variations in performance across configurations are primarily attributed to the selection of specific options.
Nonetheless, this impact is not uniform across performance indicators; an option that has a negligible effect on one indicator may significantly influence another one.
This insight empowers developers to identify high-performance configurations for specific indicators but underscores the challenge of finding configurations that excel in all indicators simultaneously.
Multi-objective optimization emerges as a promising approach to address this complex trade-off, striving to identify configurations that are near-optimal across multiple performance indicators rather than optimizing for a single indicator in isolation.

\bibliographystyle{ACM-Reference-Format}
\bibliography{jh}
\end{document}